\documentstyle[pra,aps,psfig,floats]{revtex}

\begin{document}

\renewcommand{\textfraction}{0.0}
\renewcommand{\floatpagefraction}{.7}
\setcounter{topnumber}{5}
\renewcommand{\topfraction}{1.0}
\setcounter{bottomnumber}{5}
\renewcommand{\bottomfraction}{1.0}
\setcounter{totalnumber}{5}
\setcounter{dbltopnumber}{2}
\renewcommand{\dbltopfraction}{0.9}
\renewcommand{\dblfloatpagefraction}{.7}

\draft

\twocolumn[\hsize\textwidth\columnwidth\hsize\csname@twocolumnfalse%
\endcsname

\title{Bose glass and Mott insulator phase 
in the disordered boson Hubbard model}

\author{Jens Kisker}
\address{Institut f\"ur Theoretische Physik, Universit\"at zu K\"oln,
50937 K\"oln, Germany}

\author{Heiko Rieger}
\address{HLRZ c/o Forschungszentrum J\"ulich, 52425 J\"ulich, Germany}

\date{November 29, 1996}

\maketitle

\begin{abstract}
  We study the Villain representation of the two-dimensional
  disordered boson Hubbard model via Monte Carlo simulations. It is
  shown that the probability distribution of the local susceptibility
  has a $1/\chi^2$-tail in the Bose glass phase. This gives rise to a
  divergence although particles are completely localized here
  as we prove with the help of the participation ratio. We demonstrate
  the presence of an incompressible Mott lobe within the Bose glass
  phase and show that a {\it direct} Mott-insulator to superfluid
  transition happens at the tip of the lobe. Here we find critical
  exponents $z=1$, $\nu\sim0.7$ and $\eta\sim0.1$, which are
  reminiscent of the pure three-dimensional classical XY model.
\end{abstract}


]

\newcommand{\bc}{\begin{center}}
\newcommand{\ec}{\end{center}}
\newcommand{\be}{\begin{equation}}
\newcommand{\ee}{\end{equation}}
\newcommand{\beqn}{\begin{eqnarray}}
\newcommand{\eeqn}{\end{eqnarray}}
\newcommand{\ba}{\begin{array}}
\newcommand{\ea}{\end{array}}

At zero temperature two-dimensional systems of interacting bosons can
show a quantum phase transition from an insulating phase to a
superconducting phase \cite{mpfisher,mft}. Such a transition can be
observed experimentally in granular superconductors \cite{liu} and in
$^4$He-films absorbed in arogels \cite{reppy}. By tuning a control
parameter like the disorder strength or the chemical potential the
bosons become localized in a so called Bose glass phase that is
insulating but gapless and compressible. A huge theoretical effort has
been undertaken to shed like on the universal properties of this
superconductor-to-insulator transition. In two dimensions the model
has eluded successful analytical treatment, which necessitates
numerical methods as quantum Monte Carlo simulations
\cite{scalettar,krauth,sorensen,trivedi,kawashima}, real space
renormalization group calculations \cite{singh} or strong coupling
expansion \cite{freericks}.

Apart from this generic transition the Bose glass phase itself has a
number of universal features that are relevant for experiments. Since
it is gapless various zero-frequency susceptibilities will diverge
\cite{mft}, which is reminiscent of the quantum Griffiths phase
occuring in random transverse Ising systems
\cite{griffiths,dsfisher,young,rieger,guo}, where a continuously varying
dynamical exponent parametrizes the occuring singularities. Moreover,
for weak disorder a different transition, directly from a
superconducting to a Mott-insulating phase might occur
\cite{singh,krauth}. This scenario emerges also from recent
theoretical considerations \cite{pazmandi2} and would
establish a new universality class different from the one investigated
in \cite{scalettar,sorensen,trivedi,kawashima}.

In this letter we address these two questions in a numerical approach.
We report on results obtained by extensive quantum Monte Carlo
simulations of the disordered boson Hubbard model (BH) with short
range interactions in two dimensions, which is defined by
\be
H=-t\sum_{\langle ij\rangle} (a_i^+ a_j + a_i a_j^+)
+\frac{U}{2} \sum_i n_i^2 -\sum_i\mu_i n_i
\label{BH}
\ee
where $\langle ij\rangle$ are nearest neighbor pairs on a square
lattice, $a_i^+$ ($a_i$) are boson creation (annihilation) operators,
$n_i=a_i^+a_i$ counts the number of bosons at site $i$, $U$ is the
strength of an on-site repulsion and $\mu_i$ is a random chemical
potential. We are interested in the ground state properties (i.e.\ at
temperature $T=0$) of (\ref{BH}). By using standard
manipulations \cite{sorensen} we rewrite the ground state energy
density of (\ref{BH}) as a free energy density of a classical model
\be
f = - \frac{1}{L^2 L_{\tau}} \, \ln \,
\sum_{ \{ \bf J\} }^{\nabla \cdot {\bf J}_{i,t} = 0}
\, \exp(-S\{{\bf J}\}) \; ,
\label{fe}
\ee
where the integer current variables $J_{i,t}^x$, $J_{i,t}^y$ and
$J_{i,t}^{\tau}$, live on the links of a (2+1)-dimensional cubic
lattice of linear size $L$ in the two space directions (with
coordinates $i=(x,y)$) and $L_\tau\propto1/T$ in the (imaginary) time
direction (with coordinate $t$). Ultimately one has to perform the
limit $L_\tau\to\infty$ (i.e.\ $T\to0$). The current
vector ${\bf J}_{i,t}=(J_{i,t}^x,J_{i,t}^y,J_{i,t}^{\tau})$ has to be
divergenceless on each lattice site $(i,t)$ as indicated. The
classical action $S$ is given by
\be
S\{ {\bf J} \} = 
\frac{1}{K} \sum_{(i,t)} \left\{
\frac{1}{2} {\bf J}^2_{i,t} - (\mu + v_i)J^{\tau}_{i,t}
\right\} \; .
\label{action}
\ee 
The coupling constant $K$ acts as a temperature and corresponds to
$t/U$. Note that the mapping from (\ref{BH}) to (\ref{action})
involves various approximations \cite{sorensen} and we stress right
from the beginning that we report exclusively on results for the
classical model (\ref{action}). However, as far as universal
properties are concerned, we expect them to be valid also for
(\ref{BH}). The random part $v_i$ of the local chemical potential is
distributed uniformly between $-\Delta$ and $+\Delta$. All results 
are disorder averages over at least 500 samples, obtained by Monte
Carlo simulations of the classical model (\ref{action}) with an
appropriate heat bath algorithm \cite{sorensen} at classical temperature
$K$. Details of the calculations will be published elsewhere
\cite{kisker}.

In mean field theory one expects the following $K-\mu$ phase
diagram \cite{mft}: For $\Delta<0.5$ there is a superfluid (SF) phase
at large $K$, a Bose glass (BG) phase at small $K$ and a sequence of
Mott-insulator (MI) lobes embedded into the BG phase centered around
$K=0$, $\mu$ integer. For $\Delta\ge0.5$ the Mott lobes vanish and
only the BG and SF phases remain. In this case the SF-BG transition is
generic everywhere along the phase separation line and has been
investigated extensively in \cite{sorensen} in two dimensions at the
point $K=K_c(\Delta=0.5,\mu=0.5)$. The nature of the transition at the
tip of the Mott lobes (i.e.\ at $\mu=n$ and $K=K_c'(\Delta,n)$ for
$\Delta<0.5$) is not clear and under discussion in the literature
\cite{mft,singh,scalettar,krauth,freericks}. 

Our first goal is to shed light on the Bose glass phase itself. It has
been argued \cite{mft} that here the density of states at zero
energy does not vanish, leading to a divergent superfluid
susceptibility, although the correlation length is finite. On one
hand, this behavior is reminiscent of the quantum Griffiths phase in
random transverse Ising systems \cite{dsfisher,young,rieger}. On the
other hand we demonstrate in this letter that the BG phase is
{\it different} from the Griffiths phase in the following respect:
whereas in the latter strongly coupled clusters lead to a divergence
of varying strength with varying coupling constant, essentially fully
localized excitations give rise to a uniform, logarithmic divergence
in the former. 

We study the local superfluid susceptibility, which is defined by
$\chi_i=\sum_{t=1}^{L_\tau} C_i^+(t)$ with the imaginary time
autocorrelation function $C_i^+(t)=\langle\exp\{
-\frac{1}{K}\sum_{t'=1}^t (1/2+J^{\tau}_{i,t'}-\mu_i)\}\rangle$ where
the angular brackets $\langle\cdots\rangle$ mean a thermodynamic
average.  Note that $C_i^+(\tau)$ corresponds to the local (imaginary
time) Greens function $\langle a_i(\tau)a_i^+(0)\rangle$ in the
original BH model (\ref{BH}) and the local susceptibility is simply
its (zero frequency) integral.

\begin{figure}[hbt]
\psfig{file=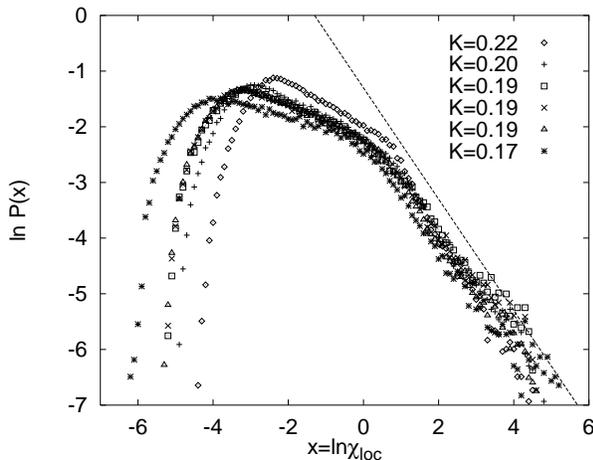,width=\columnwidth}
\caption{
  The probability distribution $P(\ln\chi_{\rm loc})$ of the local
  susceptibility 
  for various
  values of $K$ ($\Delta=0.5$, $\mu=0.5$, the BG-SF transition is at 
  $K_c=0.247$). The system size is $L=6$ and $L_\tau=200$. For
  $K=0.19$, which is deep in the Bose glass phase, also data for $L=4$ and
  $L=10$ are shown, which is indistinguishable from $L=6$.}
\label{prob_fig}
\end{figure}

The probability distribution $P(\ln\chi)$ is shown in fig.\
\ref{prob_fig} for the case $\Delta=0.5$, from which we conclude
that
\be 
\ln\,P(\ln\chi)=-\frac{d}{z}\ln\chi + const.
\label{prob}
\ee 
with $z=d=2$ throughout the BG phase. We have chosen the notation of
ref.\ \cite{rieger,young} in order to demonstrate that the dynamical
exponent $z$ that is characteristic for a
Griffiths phase \cite{griffiths} in random transverse Ising
models can also be defined in the present context and is {\it
  constant} here. Note that here $z=d$ in the BG phase
{\it and} at the critical point, although the two exponents have their
origin in different physics \cite{remark}. We also looked at weaker
disorder $\Delta=0.2$, where MI lobes are present. As soon as one
enters the latter, the distribution $P(\chi)$ is chopped off at some
characteristic value inversely proportional to the non-vanishing gap
in the MI phase. This implies furthermore that the BG phase is indeed
gapless \cite{mft}.

\begin{figure}[hbt]
\psfig{file=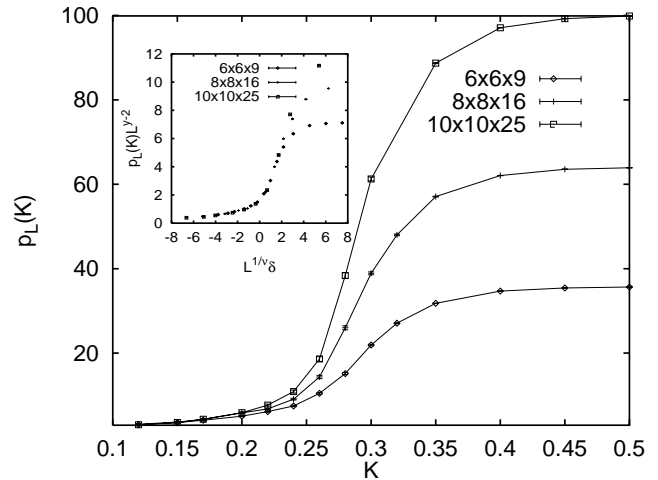,width=\columnwidth}
\caption{
  The participation ratio $p_L(K)$ (\protect{\ref{part}}) as a
  function of $K$ for various system sizes ($\Delta=0.5$, $\mu=0.5$).
  Within the Bose glass phase ($K<K_c=0.247$) $p_L$ approaches a
  constant. The insert shows a scaling plot with $\nu=0.9$ and
  $y=1.0$.  
}
\label{ratio}
\end{figure}

The relation (\ref{prob}) could be obtained by setting the
hopping matrix element $t$ to zero in (\ref{BH}), which yields a
completely {\it local} Hamiltonian. This lets us suspect that the fact
that $z$ does not vary within the BG phase is due to the local nature
of the low lying excitations. To further clarify this point we try to
quantify the degree of localization of the latter. However, since it
is not possible to obtain these excitations directly in the
representation we use, we introduce a participation ratio \cite{pazmandi}
\be
p_L=\left[\sum_{i=1}^N\langle \rho_i\rangle^2\right]_{\rm av}^{-1}
\label{part}
\ee
for the spatial density distribution
$\rho_i=L_\tau^{-1}\sum_t(J_{i,t}^{\tau,b}-J_{i,t}^{\tau,a})$ of an
{\it additional} particle, i.e.\ here we work with two replicas
$\alpha$ and $\beta$ of the system, one with fixed particle number $N$
and the other with $N+1$ ($N=L^2/2$ for the transition that
corresponds to $\mu=1/2$). $[\cdots]_{\rm av}$ denotes a disorder average. 
One expects $p_L={\cal O}(1)$ if the extra
particle is localized and $p_L={\cal O}(N)$ if it is delocalized.  The
result is shown in fig.\ \ref{ratio}, where we see very clearly, that
the additional particle becomes completely localized within the BG
phase, most probably at those sites, which allow for an extra
particle, i.e.\ $v_i\approx0$ (since we are at $\mu=1/2$). Moreover
the insert shows that $p_L(K)$ satisfies the following scaling
relation for fixed aspect ratio $L_\tau/L^z$ at the generic SF-BG
transition
\be
\frac{p_L(K)}{L^2} = L^{-y}\tilde{q}(\delta L^{1/\nu}) \;,
\ee
with $\delta=(K-K_c)/K_c$ the distance from the critical point ($K_c=0.247$),
$\nu=0.9\pm0.1$ and $z=2$ as in \cite{sorensen} and $y=1.0\pm0.1$.

\begin{figure}[hbt]
\psfig{file=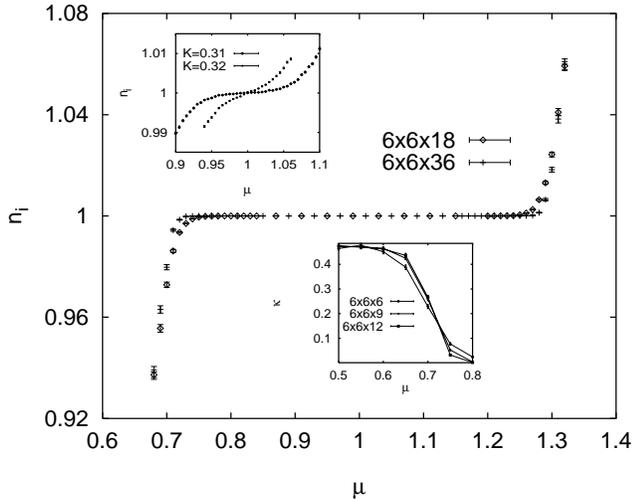,width=\columnwidth}
\caption{ The average particle number $n_i$ and compressibility
  $\kappa$ as a function of the chemical potential $\mu$
  ($\Delta=0.2$, $K=0.19$).  At $\mu=0.5$ one is deep in the Bose
  glass phase and $n_i=1/2$. The lower insert shows the
  compressibility: for large $L_\tau$ the drop in $\kappa$ becomes
  sharper.  The upper insert shows how the plateau (i.e.\ the Mott
  lobe) vanishes in the vicinity of $K=K_c'(\Delta=0.2,\mu=1)$.  }
\label{plateau}
\end{figure}

Now we consider the MI lobes within the BG phase. We choose
$\Delta=0.2$ and explore the MI-BG boundary by varying the chemical
potential between $0.5$ and $1.5$ with fixed
$K<K_c(\Delta=0.2,\mu=0.5)=0.20$. The latter point corresponds to the
generic BG-SF transition studied in \cite{sorensen}, but now with
weaker disorder. We checked that one indeed gets the same critical
exponents as for $\Delta=0.5$. In fig.\ \ref{plateau} one sees that
with increasing $\mu$ the average particle density per site increases
monotonically until it saturates in a plateau at $n_i=1$. The plateau
region indicates the boundary of the MI phase centered around $\mu=1$
with exactly one particle per site.  There is only a weak size
dependence, at least as long as one is
deep inside the BG phase, where the correlation length is very small.
The insert shows the compressibility
\be
\kappa = \frac{1}{L^d L_{\tau}} 
[\langle N^2 \rangle - \langle N \rangle^2]_{\rm av} \; ,
\label{kappa}
\ee
where $N = \sum_{(i,t)} J^{\tau}_{i,t}$ is the total number of
particles. Obviously the compressibility vanishes as soon as the
plateau, i.e.\ the MI phase, is reached. We note that we observe
extremely strong sample to sample fluctuations in the compressibility,
which necessitated an extensive disorder average (5000 samples).  With
increasing $K$ the plateau region shrinks until it vanishes at
$K_c'(\Delta=0.2,\mu=1)\approx0.325$.  This indicates the tip of the
lobe on which we focus now and for which the universality class might
be different from the generic case.

\begin{figure}[hbt]
\psfig{file=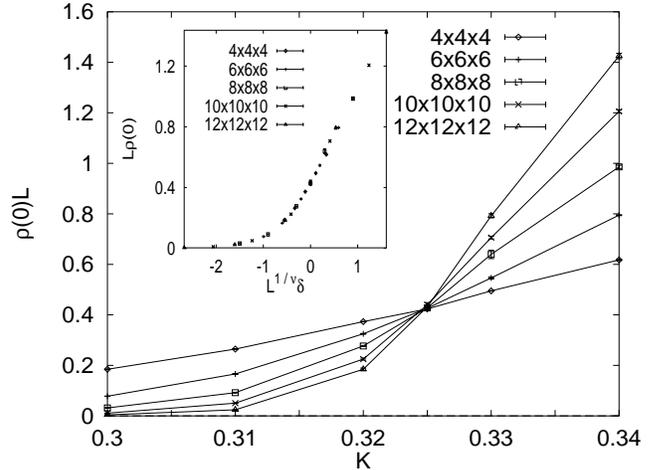,width=\columnwidth}
\caption{
  The stiffness $\rho(0)L$ at the tip of the lobe ($\Delta=0.2$,
  $\mu=1.0$). The aspect ratio is constant for $z=1$, the insert
  shows a scaling plot with $K_c=0.325$ and $\nu=0.7$.}
\label{stiffness}
\end{figure}

We fix $\mu=1$ and vary $K$. Coming from the SF-phase we first analyze
the finite size scaling behavior \cite{sorensen} of the superfluid
stiffness
\be
\rho = \frac{1}{L_{\tau}} [\langle n_x^2 \rangle ]_{av} 
\sim L^{-(d+z-2)}\,\tilde{\rho}(L^{1/ \nu}\delta,L_{\tau}/L^z)
\; ,
\label{stiff}
\ee
with $n_x = 1/L\sum_{(i,t)} J^x_{(i,t)}$ the winding number. Since we
do not know the dynamical exponent $z$ we hypothesized $z=2$ (as for
the generic case) and $z=1$ (as in the pure ($\Delta=0$) case).  For
both we performed runs with constant aspect ration $L_\tau/L^z$, and
it turned out that for $z=1$ the best data collapse could be obtained
and that only this value is also compatible with the correlation
function results discussed below. In fig.\ \ref{stiffness} we show our
results for $z=1$, where one gets a clear intersection point of
$L\rho$ at $K_c'=0.325\pm0.002$. This value is very close to the
corresponding value for the pure case $K_c'^{\rm pure}=0.333\pm0.003$
\cite{kisker} and indicates that the tip of the lobe depends very
weakly on the disorder strength or is even independent of it over some
range \cite{pazmandi2}. In the latter case the critical exponent $\nu$
might escape the inequality $\nu\ge2/d$ \cite{chayes} at this special
multicritical point \cite{pazmandi2,mefisher}, since then variation of
the disorder cannot trigger the transition as required in
\cite{chayes}. Indeed, the insert of fig.\ 4 shows a scaling plot
which yields $\nu=0.7\pm0.1<2/d=1$, which agrees well with the pure
value (see below). These results yield a consistent picture,
nevertheless we should mention that one cannot strictly exclude the
possibility that our data are not yet in the asymptotic scaling regime
and the exponent $\nu$ we estimate is only an effective exponent for
small length scales.

\begin{figure}[hbt]
\psfig{file=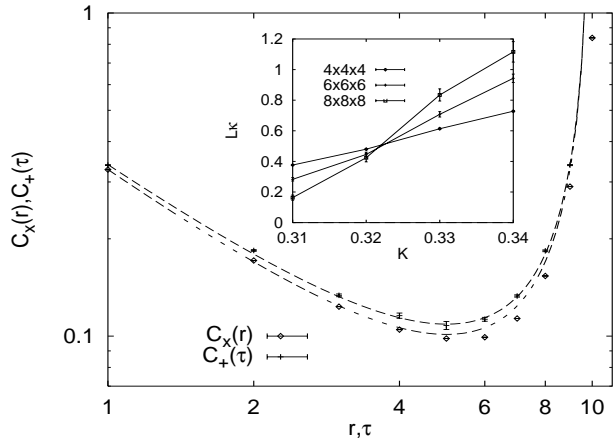,width=\columnwidth,height=6.5cm}
\caption{ The correlation functions $C_x(r)$ and $C_{+}(\tau)$ for
  system size 10x10x10 at the tip of the lobe
  ($K_c$=0.325, $\Delta$=0.2, $\mu$=1.0).  The dotted lines are least
  square fits to $C_x(r)$=$a(r^{y_x}$+$(L$--$r)^{y_x})$ and
  $C_{+}(\tau)$=$a'(\tau^{y_\tau}$+$(L_\tau$--$\tau)^{y_\tau})$, which
  gives $y_x$=--1.11 and $y_\tau$=--1.08, i.e.\
  $z$=$y_x/y_\tau$=1.0. The insert shows the compressibility
  $\kappa$ for $\Delta$=0.2, $\mu$=1.0.
  }
\label{correlation}
\end{figure}

From the data for the imaginary time correlation function $C_i(t)$ and
the spatial correlation function $C_x(r)=[\langle\exp\{
-\frac{1}{K}\sum_{x=1}^r (1/2+J^x_{i+x\hat{e}_x,t})\}\rangle]_{av}$
shown in fig.\ \ref{correlation}, we get firm support for $z=1$: The
ratio of the decay exponents $y_x$ and $y_\tau$ for $C_x$ and
$C_+(\tau)=[C_i^+(\tau)]_{\rm av}$, respectively should equal $z$ and
we find $y_x/y_\tau\approx1.0(1)$, roughly independent of how we scale
$L_\tau$ with $L$. From $y_x=d+z-2+\eta$ we get $\eta=0.1\pm0.1$.

Finally the insert of fig.\ \ref{correlation} shows the
compressibility at the above studied transition, and we observe that
it vanishes here, too. In particular the data scale well according to
$\kappa=L^{-1}\tilde{\kappa}(L^{1/\nu}\delta)$ for systems with
$L=L_\tau$, i.e.\ constant aspect ratio for $z=1$.  Hence, for {\it
  weak} disorder ($\Delta\le0.2$) we find at the tip of the lobe a
{\it direct} SF to MI transition possibly within the same universality
class as the pure model, since our estimates for $z$, $\eta$ and $\nu$
are numerically indistinguishable from the values for the the pure XY
model in 3d, which are $z=1$, $\eta=0.033(4)$ and $\nu=0.669(2)$
\cite{leguillou}.

As one can see from fig.\ \ref{stiffness} and the insert of fig.\
\ref{correlation} there is no sign of a {\it first order} transition
at the tip of the lobe, as has been suggested in \cite{freericks}.
Moreover, our conclusion disagrees with the mean-field prediction
\cite{mft} of an intervening BG phase between MI and SF phase at the
tip of the lobe for weak disorder. For stronger disorder the scenario
might change: for instance at $\Delta=0.4$ we estimate $z\approx0.4$,
which is possibly only an effective exponent and the compressibility
does not vanish immediately below the transition from the SF
phase. This indicates the existence of a threshold value for the
disorder strength: only above this threshold the mean-field prediction
might be correct \cite{singh,pazmandi2}.

To conclude, we have shown in this letter that the Bose glass phase in
the disordered boson Hubbard model and the Griffiths phase in random
transverse Ising models are closely related and that the gapless low
energy excitations are fully localized in the BG phase.  Moreover, we
presented evidence that the transition for commensurate boson
densities is directly from a Mott insulating phase to a superfluid
phase for weak disorder. The critical exponents we estimate for this
special multicritical point are different from those at the generic
BG-SG transition at incommensurate boson densities and agree with
those for the pure three-dimensional XY model. This suggests that the
latter and the tip of the lobe at weak disorder are within the same
universality class. Renormalization group and scaling arguments put
forward in \cite{pazmandi2} give strong support to this scenario.

We thank F. Pazmandi, G. T. Zimanyi and A. P. Young for intersting and
helpful discussions. This work was supported by the Deutsche
Forschungsgemeinschaft (DFG).


\vskip-0.5cm

\begin{references}

\bibitem{mpfisher}
  M. P. A. Fisher, Phys. Rev. Lett. {\bf 65}, 923 (1990).

\bibitem{mft} 
  M. P. A. Fisher, P. B. Weichman, G. Grinstein and D. S. Fisher,
  Phys. Rev. B. {\bf 40}, 546 (1988).

\bibitem{liu}
  Y. Liu et al., Phys. Rev. Lett. {\bf 67}, 2068 (1991);
  M. A. Palaanen, A. F. Hebard and R. R. Ruel,
  Phys. Rev. Lett. {\bf 69}, 1604 (1992);

\bibitem{reppy}
  P. A. Crowell, F. W. Van Keuls and J. D. Reppy,
  Phys. Rev. Lett. {\bf 75}, 1106 (1995);
  P. A. Crowell et al., Phys. Rev. B {\bf 51}, 12721 (1995).

\bibitem{scalettar} 
  R. T. Scalettar, G. G. Batrouni and G. T. Zimanyi,
  Phys. Rev. Lett. {\bf 66}, 3144 (1991).

\bibitem{krauth}
  W. Krauth, N. Trivedi and D. Ceperley,
  Phys. Rev. Lett. {\bf 67}, 2307 (1991).

\bibitem{sorensen}
  E. S. S\o rensen, M. Wallin, S. M. Girvin and A. P Young,
  Phys. Rev. Lett. {\bf 69}, 828 (1992);
  M. Wallin, E. S. S{\o}rensen, S. M. Girvin and 
  A. P. Young, Phys. Rev. B {\bf 49}, 12115 (1994).

\bibitem{trivedi}
  M. Makivi\'c, N. Trivedi and S. Ullah,
  Phys. Rev. Lett. {\bf 71}, 2307 (1993).

\bibitem{kawashima}
  S. Zhang, N. Kawashima, J. Carlson and J. E. Gubernatis,
  Phys. Rev. Lett. {\bf 74}, 1500 (1995).

\bibitem{singh}
  K. G. Singh and D. Rokshar,
  Phys. Rev. B {\bf 46}, 3002 (1992).

\bibitem{freericks} 
  J. K. Freericks and H. Monien, Phys. Rev. B {\bf 53} (1996).

\bibitem{pazmandi2}
  F. Pazmandi, G. T. Zimanyi, R. Scalettar, to be published.

\bibitem{griffiths}
  R. B. Griffiths, Phys. Rev. Lett. {\bf 23}, 17 (1969);
  B. McCoy, Phys. Rev. Lett. {\bf23}, 383 (1969).

\bibitem{dsfisher}
  D. S. Fisher, Phys. Rev. Lett. {\bf 69}, 534 (1992); 
  Phys. Rev. B {\bf 51}, 6411 (1995).

\bibitem{young} 
  A. P. Young and H. Rieger, Phys. Rev. B {\bf 53}, 8486 (1996).

\bibitem{rieger} 
  H. Rieger and A. P. Young, Phys. Rev. B {\bf 54}, 3329 (1996).

\bibitem{guo} 
  M. Guo, R. N. Bhatt and D. A. Huse,
  Phys.\ Rev.\ B {\bf 54}, 3323 (1996).

\bibitem{kisker}
  J. Kisker and H. Rieger, to be published.

\bibitem{villain} 
  J. Villain, J. Phys. {\bf 36}, 581 (1975).

\bibitem{remark} 
  The dynamical exponent $z$ that parametrizes the
  Griffiths-- or Boseglas phase as in eq.\ (\protect{\ref{prob}}) has
  its origin in purely local effects, whereas the critical exponent
  $z$ describes global, collective excitations at the critical point.

\bibitem{pazmandi}
  F. Pazmandi, G. T. Zimanyi and R. Scalettar,
  Phys. Rev. Lett. {\bf 75}, 1356 (1995).

\bibitem{chayes} 
  J. T. Chayes, L. Chayes, D. S. Fisher and T. Spencer,
  Phys. Rev. Lett. {\bf 57}, 2999 (1986).

\bibitem{mefisher} 
  M. E. Fisher and R. R. P. Singh,
  Phys. Rev. Lett. {\bf 60}, 548 (1988).

\bibitem{leguillou} 
  J. C. Le Guillou, J. Zinn-Justin, Phys. Rev. B {\bf 21}, 3976(1980).

\end{references}
\end{document}